\documentstyle[11pt,epsf]{article}
\input epsf.tex
\newcommand{\be}{\begin{equation}}
\newcommand{\ee}{\end{equation}}
\newcommand{\bea}{\begin{eqnarray}}
\newcommand{\eea}{\end{eqnarray}}
\newcommand{\nn}{\nonumber}

\begin{document}
\title{Parton Model in Lorentz Invariant Non-Commutative Space}
\author{M. Haghighat$^{1,2}$\thanks{email: mansour@cc.iut.ac.ir}\ \ \
and \ \ M. M. Ettefaghi$^{1}$\\ \\
{\it $^{1}$Department of  Physics, Isfahan University of Technology (IUT)}\\
{\it Isfahan,  Iran,} \\{\it and}\\
{\it $^{2}$Institute for Studies in Theoretical Physics and Mathematics (IPM)}\\
{\it P. O. Box: 19395-5531, Tehran, Iran.}}
\date{}
\maketitle

\begin{abstract}
We consider the Lorentz invariant non-commutative QED and complete
the Feynman rules for the theory up to the order $\theta^2$. In
the Lorentz invariant version of the non-commutative QED the
particles with fractional charges can be also considered.  We
show that in the parton model, even at the lowest order, the
Bjorken scaling violates  as $\sim\theta^2Q^4$.
\end{abstract}
\section{Introduction}
Non-commutative field theories and its phenomenological aspects
has been, recently, considered by many authors
(\cite{ncf0}-\cite{ncfp1}).  Such theories are mostly
characterized on a non-commutative space-time with  the
non-commutativity parameter $\theta_{\mu\nu}$. In the canonical
version of the non-commutative space-time one has
\begin{equation}
 \theta^{\mu\nu}=-i\left[\hat{x}^\mu,\hat{x}^\nu\right],
 \end{equation}
 where a hat indicates a non-commutative coordinate and
 $\theta_{\mu\nu}$ is a real, constant anti-symmetric matrix.
 Obviously, the constant vectors $\theta_{0i}$ and $\theta_{ij}$
 imply preferred directions in a given Lorentz frame which leads
 to violation of the Lorentz symmetry.  Since the Lorentz symmetry is an almost exact
 symmetry of nature, it is natural to explore the non-commutative (NC) field
 theories that are Lorentz invariant from the beginning. In this
 new class of NC theories the parameter of non-commutativity is
 not a constant but is an operator which transforms as a Lorentz
 tensor (\cite{lcncf0}-\cite{lcncf1}). Of course in this way one needs to generalize the star
 product and operator trace for functions of both $x^\mu$ and
 $\theta_{\mu\nu}$, appropriately.  However, in both cases
 experiment should confirm the theories.  The obtained upper
 bounds
 for the violating Lorentz non-commutative field theory are two
 folds: the first one comes from bound states such as the Hydrogen
 atom or the positronium (\cite{ncf4}, \cite{ncfp0}) and the second one
 is obtained by scattering
 processes for example the electron-electron and the electron-photon scattering
 and so on (\cite{ncf3}, \cite{ncfp1}), see Table 1.
\begin{table}[h]
\begin{center}
\begin{tabular}{|c|c|c|}\hline
NC-parameter &bound state&scattering \\ \hline $\Lambda_{NC}$ &$\sim$100 GeV&0.1-2.5 TeV\\
\hline $\Lambda_{LCNC}$&?&0.1-1 TeV\\ \hline
\end{tabular}
\caption{ Upper bounds on the non-commutativity parameter.
$\Lambda_{NC}=1/\sqrt{\Theta}$ describes the parameter of
non-commutativity in the standard NC-space while $\Lambda_{LCNC}$
shows NC-parameter for LCNC-space.}
\end{center}
\end{table}
 In the case of Lorentz conserving non-commutative (LCNC) field theory
  scattering process is only investigated.  The dimensional
quantity $\theta$ in the non-commutative space imply new aspects
in the parton model as well. In this paper we
 explore parton model in the lowest order in which a virtual photon
 interacts with partons inside a nucleon in a LCNC-space.
 For this purpose one should consider LCNCQED to find the effect
 of non-commutativity on the form factors.\\
 In section 2 we introduce Feynman rules for LCNCQED. In section 3
 we study the parton model in the non-commutative space in the lowest
 order and show how the form factors in the electron-nucleon scattering
 depend on $Q^2$ and the Bjorken
 scaling is violated. Finally, we compare our results with the
 experimental data and give an upper bound on the parameter of
 non-commutativity.

\section{Lorentz conserving NCQED}
To construct the non-commutative field theories that are Lorentz
invariant one needs to generalize the parameter of
non-commutativity.  Now we review the formalism of the Lorentz
conserving NCQED introduced by Carlson, Carone and Zobin (CCZ)
\cite{lcncf0}. In the CCZ approach of NCQED
$\hat{\theta^{\mu\nu}}$ is an operator and satisfies the
following algebra:
\begin{equation}
\begin{array}{ll}
&[\hat{x}^\mu,\hat{x}^\nu]=i\hat{\theta^{\mu\nu}},\\
&[\hat{\theta}^{\mu\nu},\hat{x}^\lambda]=0,\\
&[\hat{\theta}^{\mu\nu},\hat{\theta}^{\alpha\beta}]=0, \label{a1}
\end{array}
\end{equation}

where $\hat{\theta}^{\mu\nu}$ is antisymmetric tensor that is not
constant but transforms as a Lorentz tensor.  The action for
field theories on noncommutative spaces is then obtained using
the Weyl-Moyal correspondence, according to that, in order to
find the noncommutative action, the usual product of fields
should be replaced by the star product:
\begin{equation}
 f\ast
g(x,\theta)=f(x,\theta)\exp(\frac{1}{2}\overleftarrow{\partial}_\mu
\theta^{\mu\nu}\overrightarrow
{\partial}_\nu)g(x,\theta). \label{a3}
\end{equation}
It should be noted that here the mapping to c-number coordinates
involve ${\theta}^{\mu\nu}$ as a c-number due to the presence of
the operator $\hat{\theta}^{\mu\nu}$ in the Lorentz-conserving
case. In this formulation, the operator trace that is a map from
operator space to numbers, is defined as
\begin{equation}
Tr\hat{f}=\int d^4xd^6\theta W(\theta)f(x,\theta), \label{a4}
\end{equation}
where $W(\theta)$ is a Lorentz invariant weight function and is
assumed to be positive and even function of ${\theta}$ therefore
one has
\begin{equation}
\int d^6\theta W(\theta)\theta^{\mu\nu}=0. \label{a5}
\end{equation}
Furthermore, the weight function is assumed to fall sufficiently
fast so that all integrals are well defined. Now the properties
of $W(\theta)$ and the definition of the operator trace
Eq.(\ref{a4}) allows one to extract the interactions in the
Lorentz conserving noncommutative field theory.  To this end the
action can be written as follows
\begin{equation}
s=\int d^4xd^6\theta W(\theta)\cal{L}(\phi,\partial\phi)_*,
\label{a6}
\end{equation}
where $\cal{L}(\phi,\partial\phi)_*$ depends on both $x$ and
$\theta$ and its subscript indicates the $*$-product which is
defined in Eq.(\ref{a3}).  For a $U(1)$ gauge theory the gauge
invariant Lagrangian is
\begin{equation}
{\cal{L}}=\int d^6\theta
W(\theta)[-{\frac{1}{4}}F_{\mu\nu}*F^{\mu\nu}+\bar{\psi}*(iD\!\!\!\!/-m)*\psi
], \label{a61}
\end{equation}
where $\psi$ is a matter field with charge $q$ and for a gauge
field $A$
\begin{equation}
D_{\mu}=\partial_\mu+iqA_\mu, \label{a62}
\end{equation}
and
\begin{equation}
F_{\mu\nu}=\partial_\mu A_\nu-\partial_\nu A_\mu
+iq[A_\mu,A_\nu]_*. \label{a63}
\end{equation}
The matter field and the gauge field as well as the parameter of
the gauge transformation $\Lambda(x,\theta) $ are functions of
both $x$ and $\theta$.  Therefore, using the same method as
applied in the construction of $SU(N)$ noncommutative gauge
theories \cite{sun}, one should expand the fields as a power
series in the variable $\theta$ as follows
\bea
\Lambda_\alpha(x,\theta)&=&\alpha(x)+\theta^{\mu\nu}\Lambda^{(1)}_{\mu\nu}(x_i,\alpha)
    +\theta^{\mu\nu}\theta^{\eta\delta}\Lambda^{(2)}_{\mu\nu\eta\delta}(x,\alpha)+...,\nn\\
A_\rho(x,\theta)&=&A_\rho(x)+\theta^{\mu\nu}A^{(1)}_{\mu\nu\rho}(x_i)
    +\theta^{\mu\nu}\theta^{\eta\delta}A^{(2)}_{\mu\nu\eta\delta}(x)+...,\\
\psi(x,\theta)&=&\psi(x)+\theta^{\mu\nu}\psi^{(1)}_{\mu\nu}(x_i,\alpha)
    +\theta^{\mu\nu}\theta^{\eta\delta}\psi^{(2)}_{\mu\nu\eta\delta}(x)+...,\nn
\label{a7}
 \eea
 where $\alpha(x)$, $A(x)$ and $\psi(x)$ are
gauge parameter, gauge field and matter field in the commutative
space, respectively, and the coefficients of $\theta_{\mu\nu}$
can be obtained as
\begin{eqnarray}
\begin{array}{ll}
&\Lambda^{(1)}_{\mu\nu}(x,\alpha)=\frac{-q}{2}\partial_\mu\alpha(x)A_\mu(x),\\
&\Lambda^{(2)}_{\mu\nu\eta\delta}(x,\alpha)=-\frac{q^2}{2}\partial_\mu\alpha(x)A_\eta(x)
\partial_\delta A_\nu(x),\\
&A^{(1)}_{\mu\nu\rho}(x)=\frac{q}{2}A_\mu(\partial_\nu A_\rho+{\cal F}^0_{\nu\rho}),\\
&A^{(2)}_{\mu\nu\eta\delta\rho}(x)=\frac{q^2}{2}(A_\mu
A_\eta\partial_\delta{\cal F}^0_{\nu\rho} -\partial_\nu
A_\rho\partial_\eta A_\mu A_\delta
+A_\mu{\cal F}^0_{\nu\eta}{\cal F}^0_{\delta\rho}),\\
&\psi^{(1)}_{\mu\nu}(x)=\frac{q}{2}A_\mu\partial_\nu\psi,\\
&\psi^{(2)}_{\mu\nu\eta\delta}(x)=-\frac{q}{8}(-i\partial_\mu
A_\eta\partial_\nu
\partial_\delta\psi-qA_\mu A_\eta\partial_\nu\partial_\delta\psi-2qA_\mu\partial_\nu A_\eta
\partial_\delta\psi-qA_\mu{\cal F}^0_{\nu\eta}\partial_\delta\psi\\
&+\frac{q}{2}\partial_\mu A_\eta\partial_\nu A_\delta\psi
+iq^2A_\mu A_\delta\partial_\eta A_\nu\psi), \\
\label{a8}
\end{array}
\end{eqnarray}
where ${\cal{F}}^0_{\mu\nu}=\partial_\mu A_\nu-\partial_\nu
A_\mu$. Now we need ${\cal{L}}(x)$, to extract the interactions,
that can be obtained by inserting the above relations into
Eq.(\ref{a61}) and performing the integration on $\theta$ using
the weighted average
\begin{equation}
\int d^6\theta W(\theta)\theta^{\mu\nu}\theta^{\eta\rho}=
\frac{\langle\theta^2\rangle}{12}(g^{\mu\eta}g^{\nu\rho}-g^{\mu\rho}g^{\eta\nu}),
\label{a9}
\end{equation}
where
\begin{equation}
\langle\theta^2\rangle=\int d^6\theta
W(\theta)\theta^{\mu\nu}\theta_{\mu\nu}.\label{a10}
\end{equation}
Therefore up to the order ${\theta}^2$ the second term in the
Lagrangian density Eq.(\ref{a61}) can be rewritten as follows
\begin{equation}
\bar{\psi}*(iD\!\!\!\!/-m)*\psi={\cal{L}}_0+{\cal{L}}_{qq\gamma}+{\cal{L}}_{qq\gamma\gamma}
+{\cal{L}}_{qq\gamma\gamma\gamma},
\end{equation}
where
\begin{eqnarray}
\begin{array}{ll}
&{\cal{L}}_0=\bar{\psi}^{(0)}(i\partial\!\!\!/-m){\psi}^{(0)},\\
&{\cal{L}}_{qq\gamma}=-q(\bar{\psi}^{(0)}*{A\!\!\!/}^{(0)}){\psi}^{(0)}+
\bar{\psi}^{(0)}(i\partial\!\!\!/-m){\psi}^{(2)}+
\bar{\psi}^{(2)}(i\partial\!\!\!/-m){\psi}^{(0)},\\
&{\cal{L}}_{qq\gamma\gamma}=-q(\bar{\psi}^{(0)}*{A\!\!\!/}^{(0)}){\psi}^{(1)}-
q(\bar{\psi}^{(1)}*{A\!\!\!/}^{(0)}){\psi}^{(0)}-
q\bar{\psi}^{(0)}{A\!\!\!/}^{(0)}{\psi}^{(2)}\\
&-q\bar{\psi}^{(2)}{A\!\!\!/}^{(0)}{\psi}^{(0)}-q
(\bar{\psi}^{(0)}*{A\!\!\!/}^{(1)}){\psi}^{(0)} +
\bar{\psi}^{(1)}(i\partial\!\!\!/-m){\psi}^{(1)},\\
&{\cal{L}}_{qq\gamma\gamma\gamma}=-q\bar{\psi}^{(0)}{A\!\!\!/}^{(1)}{\psi}^{(1)}
-q\bar{\psi}^{(1)}{A\!\!\!/}^{(1)}{\psi}^{(0)} -q
\bar{\psi}^{(0)}{A\!\!\!/}^{(2)}{\psi}^{(0)}-q
\bar{\psi}^{(1)}{A\!\!\!/}^{(0)}{\psi}^{(1)}. \label{lagran}
\end{array}
\end{eqnarray}
The Feynman rules can be extracted from Eq.(\ref{a61}), for
example ${\cal{L}}_0$ in Eq.(\ref{lagran}) leads to the ordinary
propagator for a fermion field while the rest terms contain a
variety of vertices. Comparing Eqs.(\ref{lagran}) and (\ref{a8})
we see that, up to the order ${\theta}^2$, ${\cal{L}}_{qq\gamma}$
contributes to 2-fermion-1-photon ($qq\gamma$), 2-fermion-2-photon
($qq\gamma\gamma$) and 2-fermion-3-photon
($qq\gamma\gamma\gamma$) vertices while
${\cal{L}}_{qq\gamma\gamma\gamma}$ has just contribution on
$qq\gamma\gamma\gamma$-vertex and ${\cal{L}}_{qq\gamma\gamma}$
has terms which contains two fermions and two, three and four
photons.  In \cite{lcncfp0} the Feynman rules for
$qq\gamma$-vertex in general case and $qq\gamma\gamma$-vertex
when all fermions and photons are on the mass shell is obtained.
Here we complete the Feynman rules for the interaction between
fermions and photons
up to the order $\theta^2$ as follows:\\

$qq\gamma$-vertex in general case \cite{lcncfp0}:\\
$$-iq[\gamma_{\mu}+\frac{\langle\theta^2\rangle}{96}\{({p\!\!\!/}_1-m)p^2_2p^{\mu}_3-
({p\!\!\!/}_2-m)p^2_1p^{\mu}_3+({p\!\!\!/}_2-m)p_1\cdot
p_3p^{\mu}_1$$
$$-({p\!\!\!/}_1-m)p_2\cdot p_3p^{\mu}_2 +
\frac{1}{2}((p_1\cdot p_3)^2 + (p_2\cdot p_3)^2)\gamma^{\mu}\}].
$$

${qq\gamma\gamma}$-vertex with all fermions and photons are on the mass shell \cite{lcncfp0}:\\
$$iq^2\frac{\langle\theta^2\rangle}{96}\{p_1\cdot p_3[p^{\rho}_2\gamma^{\eta}
-p^{\eta}_1\gamma^{\rho}]+ p_1\cdot p_4[p^{\eta}_2\gamma^{\rho}
-p^{\rho}_1\gamma^{\eta}]+
({p\!\!\!/}_3-{p\!\!\!/}_4)[p^{\rho}_1p^{\eta}_2 -
p^{\eta}_1p^{\rho}_2\}.$$

\begin{figure}
\centerline{\epsfxsize=2in\epsffile{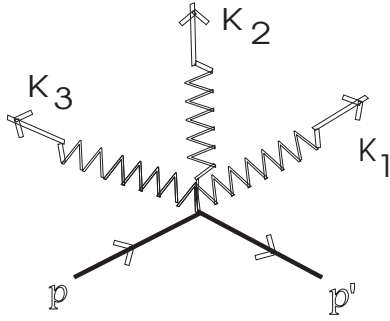}}
\caption{${qq\gamma\gamma\gamma}$-vertex with all fermions and
photons on mass shell.  } \label{3photon}
\end{figure}

${qq\gamma\gamma\gamma}$-vertex with all fermions and photons on shell:\\
\begin{eqnarray}
\begin{array}{ll}
\frac{q^3\langle\theta^2\rangle}{96}[&(k_1+k_2)^{\tau}({k\!\!\!/}_1+{k\!\!\!/}_2)
g^{\sigma\rho} + (k_1+k_3)^{\sigma}({k\!\!\!/}_1+{k\!\!\!/}_3)
g^{\tau\rho}+ (k_2+k_3)^{\rho}({k\!\!\!/}_2+{k\!\!\!/}_3)
g^{\tau\sigma}\\ &+(k_2+k_1)\cdot k_3\gamma^{\tau}g^{\sigma\rho}
+ (k_3+k_1)\cdot k_3\gamma^{\sigma}g^{\tau\rho} +(k_3+k_2)\cdot
k_1\gamma^{\rho}g^{\tau\sigma}],\label{2q3g}
\end{array}
\end{eqnarray}
where $k_i, i=1, 2, 3,$ are defined in Fig. (\ref{3photon}).\\

${qq\gamma\gamma\gamma\gamma}$-vertex in general case is zero
because in Eq.(\ref{lagran}) terms which contain two fermions and
four photons are $-q\bar{\psi}^{(0)}{A\!\!\!/}^{(0)}{\psi}^{(2)}$
and $-q\bar{\psi}^{(2)}{A\!\!\!/}^{(0)}{\psi}^{(0)}$ or
$$\frac{iq^4}{8}(\bar{\psi}^{(0)}A{\!\!\!/}A_{\mu}A_{{\nu}^\prime}\partial_{{\mu}^\prime}A_\nu
\psi^{(0)}-\bar{\psi}^{(0)}A_{\mu}A_{{\nu}^\prime}\partial_{{\mu}^\prime}A_\nu
A{\!\!\!/}\psi^{(0)}),$$ which is equal to zero.  For the pure
gauge vertices up to the order $\theta^2$ there is only four
photon vertex which has been already obtained in \cite{lcncf0}.
These rules for vertices is relevant to study the phenomenology
of LCNCQED.  For instance at the lowest order (i.e. tree level)
the correction to $qq\gamma$-vertex can be obtained for on shell
fermions as
\begin{equation}
-iq\{1+\frac{\langle\theta^2\rangle}{384}k^4\}\gamma^\mu,
\label{a12}
\end{equation}
where $k$ is the photon momentum.  For the cross section of the
process $e^-e^+\longrightarrow\mu^+\mu^-$ such a correction
results in
\begin{equation}
\frac{d\sigma}{d\cos\theta}=(\frac{d\sigma}{d\cos\theta})_{QED}
({1+\frac{\langle\theta^2\rangle}{96}s^2}), \label{a13}
\end{equation}
and for the cross process $e^-\mu^-\longrightarrow e^-\mu^-$  $s$
should be replaced by $t$ in Eq.(\ref{a13}).
\section{Parton model at the lowest order in LCNC-space }
In the canonical version of NCQED the matter fields with charges
$0$ or $\pm 1$ are allowed i.e. charged leptons and photon. But in
the LCNCQED the quarks as well as the leptons and photon, can be
accommodated in the theory.  Therefore we can examine the NC
effects for the processes which contain quarks. For this purpose
we consider inclusive inelastic electron-nucleon  scattering.  In
this process the electron, at the lowest order of the parton
model, interacts  with free charged partons via one photon
exchange therefore modification in the obtained results with
respect to the usual space can be expected. To this end we
explore the differential cross section for the unpolarized e-N
scattering as follows
\begin{equation}
\frac{d^2\sigma}{dE^\prime
d\Omega}|_{eN}=\frac{E^\prime\alpha^2}{EQ^4}L^{\mu\nu}W_{\mu\nu},\label{a231}
\end{equation}
where $E$, $E^\prime$, $\sqrt{-Q^2}$, $L^{\mu\nu}$ and
$W_{\mu\nu}$ are initial and final energy of electron, the
momentum transfer, the electron and the nucleon scattering
tensor, respectively. The $ee\gamma$ vertex in the LCNCQED is
given in Eq.(\ref{a12}) therefore $L^{\mu\nu}$ can be easily
obtained as
\begin{eqnarray}
\begin{array}{ll}
L_{\mu\nu}&=\frac{1}{2}(1+\frac{\langle\theta^2\rangle}{384}Q^4)^2tr((P{\!\!\!\!/}+m)
\gamma_\mu(P{\!\!\!\!/}+m)\gamma_\nu)\\
          &=2(1+\frac{\langle\theta^2\rangle}{384}Q^4)^2(p_\mu p^\prime_\nu+p_\nu p^\prime_\mu
-g_{\mu\nu}p.p^\prime+g_{\mu\nu}m^2). \label{a23}
\end{array}
\end{eqnarray}
The inelastic nucleon scattering tensor is proportional to the
absolute square of the nucleonic current therefore we need the
vertex function ($\Gamma_\mu$) for the nucleonic current in the
NC space.  Since it is a Lorentz vector therefore the most general
form of $\Gamma_\mu$ can be written as
\begin{equation}
\Gamma_\mu=A\gamma_\mu+BP^\prime_\mu+CP_\mu+iDP^{\prime\nu}\sigma_{\mu\nu}
+iEP^{\prime\nu}\sigma_{\mu\nu}+FP^\nu\theta_{\mu\nu}+GP^{\prime\nu}\theta_{\mu\nu},
\label{a14}
\end{equation}
where A,B,...,G depend on the Lorentz invariant quantity. The
gauge invariance and using the Gordon identity leads to
\begin{equation}
\bar{u}(P^\prime)\Gamma_\mu(P^\prime,P)u(P)=\bar{u}(P^\prime)[A\gamma_\mu+iB
(P^\prime-P)^\nu\sigma_{\mu\nu}+C(P-P^\prime)^\nu\theta_{\mu\nu}]u(p).
\label{a17}
\end{equation}
We can now construct $W_{\mu\nu}$ as follows:
\begin{eqnarray}
\begin{array}{lll}
W_{\mu\nu}&=&\frac{1}{2}\sum_{spin}[\bar{u}(P^\prime)\Gamma_\mu
u(P)]^*[\bar{u}(P^\prime)
\Gamma_\nu u(P)]\\
          &=&\frac{1}{2}tr\{(A\gamma_\mu- i B(P-P^\prime)^\lambda\sigma_{\mu\lambda}+C(P-P^\prime)^\lambda
\hat{\theta}_{\mu\lambda}(P^\prime{\!\!\!\!\!/}+M))\\
          &
          &(A\gamma_\nu+iB(P-P^\prime)^\rho\sigma_{\nu\rho}+C(P-P^\prime)^\rho\hat{\theta}_{\nu\rho}(P{\!\!\!\!/}+M))\}.
\label{a18}
\end{array}
\end{eqnarray}
Since the weight function is an even function of
$\theta^{\mu\nu}$ the odd functions of $\theta^{\mu\nu}$ have not
any contribution on the cross section thus Eq.(\ref{a18}) can be
cast into
\begin{equation}
W_{\mu\nu}=W^0_{\mu\nu}+4P.P^\prime
\frac{\langle\theta^2\rangle}{12}C(q^2g_{\mu\nu}- q_\mu
q_\nu),\label{a200}
\end{equation}
where $q_\mu=(P-P^\prime)_\mu$, $q^2=-Q^2$ and $W^0_{\mu\nu}$ is
the commutative counterpart of the scattering tensor and is given
as
\begin{equation}
W^0_{\mu\nu}=(-g_{\mu\nu}+\frac{q_\mu
q_\nu}{q^2})W_1+(P_\mu-q_\mu\frac{P.q}{q^2})
(P_\nu-q_\nu\frac{P.q}{q^2})\frac{W_2}{M_N^2}. \label{a20}
\end{equation}
$W_1$ and $W_2$ are the structure functions and depend on the
Lorentz invariant quantity such as $Q^2$, $\nu=P.Q$ and
$\langle\theta^2\rangle$. One can see that the second term in
Eq.(\ref{a200}) can be absorbed in $W_1$ and  $W_{\mu\nu}$ can be
generally written as
\begin{equation}
W^{inel}_{\mu\nu}=(-g_{\mu\nu}+\frac{q_\mu
q_\nu}{q^2})W_1(Q^2,\nu,\langle\theta^2\rangle)+
(P_\mu-q_\mu\frac{P.q}{q^2})
(P_\nu-q_\nu\frac{P.q}{q^2})\frac{W_2(Q^2,\nu,\langle\theta^2\rangle)}{M_N^2},
\label{a22}
\end{equation}
where
\begin{equation}
W_1(Q^2,\nu,\langle\theta^2\rangle)= W_1+4CP.P^\prime
\frac{\langle\theta^2\rangle}{12}Q^2\,\,\,\,
,\,\,\,\,W_2(Q^2,\nu,\langle\theta^2\rangle)= W_2. \label{a221}
\end{equation}
 In the high energy limit the mass of the electron can be
neglected and the differential cross section for the inclusive
e-N scattering, in terms of Bjorken variable $x= \frac{Q^2}{2\nu}$
and the inelasticity parameter $y=\frac{E-E^\prime}{E}$, can be
cast into
\begin{eqnarray}
\begin{array}{ll}
\frac{d^2\sigma}{dxdy}|_{eN}&=\frac{4\pi
s\alpha^2}{Q^4}(1+\frac{\langle\theta^2\rangle}{384}
Q^4)^2[xy^2F_1^{eN}(Q^2,x,\langle\theta^2\rangle)\\
          &+(1-y-\frac{xyM^2_N}{s})F_2^{eN}
(Q^2,x,\langle\theta^2\rangle)], \label{a24}
\end{array}
\end{eqnarray}
where $M_N$ is the nucleon mass, $F_1^{eN}=M_NW_1^{eN}$,
$F_2^{eN}=\frac{\nu}{M_N}W_2^{eN}$ and $s$ is the mandelstam
variable. In the parton model at the  lowest order, one consider
the elastic scattering of the electron off a free point charged
parton with mass $M_i$, momentum $P_i$ and charge $q_ie$.
Therefore the cross section for this scattering can be easily
constructed from the results for the electron-muon scattering
(see Eq.(\ref{a13})) as
\begin{equation}
\frac{d\sigma_i}{dQ^2}=(\frac{d\sigma_i}{dQ^2})_{QED}(1+\frac{\langle\theta^2\rangle
t_i^2}{96}), \label{a25}
\end{equation}
where $t_i$ is the Mandelstam variable for the parton $i$. If we
neglect the electron and the partons masses in the Briet frame of
reference
\begin{eqnarray}
\begin{array}{ll}
&q_\mu=(0,0,0,\sqrt{-q^2})=(0,0,0,\sqrt{Q^2})\\
&P_\mu=(\tilde{P},0,0,-\tilde{P})\hspace{1cm},\hspace{1cm}\tilde{P}\gg
M_N\hspace{1cm},
\hspace{1cm}P^2=\tilde{P}^2-\tilde{P}^2=0\approx M_N^2,\\
\label{a26}
\end{array}
\end{eqnarray}
and after a little algebra the cross section in terms of $x$ and
$y$ variables becomes
\begin{equation}
\frac{d^2\sigma_i}{dxdy}=\frac{4\pi\alpha^2q_i^2x}{Q^2}(\frac{s^2+u^2}{2s^2})\delta(\xi_i-x)(1+
\frac{\langle\theta^2\rangle t^2}{96}), \label{a29}
\end{equation}
where $\xi_i$ is a fraction of the nucleon's total momentum
carried by the $i$th parton (i.e. $P_i^\mu=\xi_iP^\mu$) and the
Mandelstam variables for $i$th parton in terms of the Mandelstam
variables for the whole nucleon are
\begin{eqnarray}
\begin{array}{ll}
&s_i=(p+P_i)^2=\xi_is,\\
&t_i=(P_i-P_i^\prime)^2=t=-Q^2,\\
&u_i=(p^\prime-P_i)^2=\xi_iu. \label{a28}
\end{array}
\end{eqnarray}
Various types of partons carry a different fraction of the parent
nucleon's momentum therefore for the parton momentum distribution
function $f_i(\xi_i)$ with the appropriate normalization:
\begin{equation}
\int_0^1d\xi_if_i(\xi_i)=1,
\end{equation}
one has
\begin{equation}
\frac{d^2\sigma}{dxdy}=\frac{2\pi\alpha^2s}{Q^4}[(y-1)^2+1]\sum_if_i(x)q_i^2x(1+
\frac{\langle\theta^2\rangle t^2}{96}). \label{a30}
\end{equation}
Now comparing Eq.(\ref{a30}) with Eq.(\ref{a24}), where $M_N=0$,
reads
\begin{eqnarray}
\begin{array}{lll}
&2xy^2F_1^{eN}(Q^2,x,&\!\!\!\!\langle\theta^2\rangle)+
2(1-y)F_2^{eN}(Q^2,x,\langle\theta^2\rangle)=\\&
&(y^2+2(1-y))\sum_if_i(x)q_i^2x(1+\frac{\langle\theta^2\rangle
Q^4}{192} ). \label{a31}
\end{array}
\end{eqnarray}
Equation (\ref{a31}) shows that
\begin{equation}
F_2^{eN}(Q^2,x,\langle\theta^2\rangle)=\sum_if_i(x)q_i^2x(1+\frac{\langle\theta^2\rangle
Q^4}{192} ),\label{a32}
\end{equation}
in other words the parton model at the lowest order in LCNCQED
violates the Bjorken scaling but Callan-Gross relation still
holds:
\begin{equation}
F_2^{eN}(Q^2,x,\langle\theta^2\rangle)=2xF_1^{eN}(Q^2,x,\langle\theta^2\rangle).
\label{a33}
\end{equation}
One can use the data on the deep inelastic e-N scattering to
obtain the upper bound on the value of the parameter of
non-commutativity in the LCNCQED, $\Lambda_{LCNC}$. Equation
(\ref{a32}) shows that
\begin{equation}
{\frac{F_2^{NC}-F_2^{0}}{F_2^{0}}}= \frac{\langle\theta^2\rangle
Q^4}{192}=\frac{Q^4}{16\Lambda^4_{LCNC}}, \label{a34}
\end{equation}
where
$\Lambda_{LCNC}=(\frac{12}{\langle\theta^2\rangle})^{\frac{1}{4}}$.
The measurements of $F_2$ structure function in deep inelastic
scattering can provide a test on the non-commutativity of space.
For example $F_2$ in positron-proton neutral current scattering
has been measured with statistical and systematic uncertainties
below $2\%$ \cite{expF2}.  Therefore, as an estimation, one
percent error in the experimental value of the structure function
for $\sqrt{Q^2}=200 GeV$ results in $\Lambda_{LCNC}\sim 300 GeV$.
\section*{Summary}
We completed the Feynman rules for the Lorentz conserving
non-commutative QED up to the order $\langle\theta^2\rangle$.
Besides two fermions and one and two photons vertices which has
already introduced in \cite{lcncfp0} there is a two fermions and
three photon vertex given in Eq.(\ref{2q3g}).  The parameter of
non-commutativity is a dimensional quantity therefore the
dimensionless form factors in lepton-nucleon scattering should
depend on $\langle\theta^2\rangle Q^4$ and violate the  Bjorken
scaling. We explicitly obtained this violation in Eqs.(\ref{a32})
and (\ref{a34}) while the Callan-Gross relation still holds as is
shown in Eq.(\ref{a33}).  The obtained results provide an
experimental tool to find if the nature can be described by
LCNC-space or not, for $Q\sim 1.1\Lambda_{LCNC}- 2\Lambda_{LCNC}$
there is $10\%-100\%$ correction which can be easily verified in
comparison with the experimental data.
\section*{Acknowledgment}
 The financial support of IUT research council and
IPM are acknowledged.

\end{document}